\documentclass[twocolumn]{aastex62}
\usepackage{graphics,epsf}
\usepackage{amsmath}                
\usepackage{amsfonts}               
\usepackage{amssymb}                
\usepackage{epsfig}                 
\usepackage{appendix}
\usepackage{graphicx}
\usepackage{float}
\usepackage{color}
\usepackage[para,online,flushleft]{threeparttable}

\newcommand{\cm}{{~\rm cm}}

\newcommand{\km}{{~\rm km}}
\newcommand{\s}{{~\rm s}}

\newcommand{\erg}{{~\rm erg}}
\newcommand{\yr}{{~\rm yr}}

\newcommand{\Gyr}{{~\rm Gyr}}

\newcommand{\days}{{~\rm days}}

\begin{document}

\title{Supernovae Ia in 2019 (review): a rising demand for spherical explosions}

\email{soker@physics.technion.ac.il}

\author{Noam Soker}
\affiliation{Department of Physics, Technion – Israel Institute of Technology, Haifa 3200003, Israel}
\affiliation{Guangdong Technion Israel Institute of Technology, Guangdong Province, Shantou 515069, China}

\begin{abstract}
I review new studies of type Ia supernovae (SNe Ia) from 2019, and use these to improve the comparison between the five binary SN Ia scenarios. 
New low polarisation measurements solidify the claim that most SN Ia explosions are globally spherically symmetric (clumps are possible). Explosions by dynamical processes, like explosions that take place during a merger process of two white dwarfs (WDs) in the double degenerate (DD) scenario, or during an accretion process in the double detonation (DDet) scenario and in the single degenerate (SD) scenario, lead to non-spherical explosions, in contradiction with observations of normal SNe Ia. I argue that these (DD, DDet, SD) scenarios account mainly for peculiar SNe Ia. The explosion of a Chandrasekhar mass ($M_{\rm CH}$) WD (deflagration to detonation process) has a global spherical structure that is compatible with observations.   
To reach spherical explosions, SN Ia scenarios should allow for a time delay between the formation of an $M_{\rm Ch}$-WD and its explosion. As such, I split the DD scenario to a channel without merger to explosion delay (MED) time (that forms mainly peculiar SNe Ia), and a channel with a MED, the DD-MED channel (scenario). I speculate that the main contributors to normal SNe Ia are the core degenerate (CD) scenario, the DD-MED scenario, both have $M_{\rm CH}$ spherical explosions, and the DD scenario that has sub-$M_{\rm CH}$ non-spherical explosions. 
\end{abstract}



\section{INTRODUCTION}
\label{sec:intro}
     
The year 2019 saw the continuation of the debate between the supporters of the five binary scenarios for type Ia supernovae (SNe Ia; in \citealt{Soker2019CEE} I justify this partition to five scenarios). Despite several scientific meetings and tens of papers (e.g., \citealt{LiuStancliffe2018, Ablimitaeda2019, Ashalletal2019b, Bravoetal2019, Brownetal2019, Calderetal2019, Chakradharietal2019, Chenetal2019Xingzhuo, Graur2019, Grauretal2019NatAs, Heringeretal2019b, KonyvesTothetal2019, Liuetal2019, Milesetal2019, Moriyaetal2019, Siebertetal2019, Taubenbergeretal2019, Wuetal2019, Meng2019,  Yaoetal2019, DiStefano2020, HanXetal2020, KushnirKat2020a, KushnirKat2020b, LeungNomoto2020, Mazzalietal2020J, Milleretal2020, Panetal2020, Pellegrinoetal2020, Shinglesetal2020}), the community did not come closer to a consensus. A prominent example is the existence of studies that argue for Chandrasekhar SN Ia explosion alongside studies that argue for sub-Chandrasekhar SNe Ia, and studies that argue for both (e.g., \citealt{LevanonSoker2019b, Sarbadhicaryetal2019, Seitenzahletal2019}). 
However, I do find some significant results in studies from 2019 that constrain some scenarios and some channels within scenarios. 
Since there are several reviews of SNe Ia from the years 2018-2019 (\citealt{LivioMazzali2018, Wang2018, Soker2018Rev, Jhaetal2019NatAs, RuizLapuente2019}; older reviews include, e.g., \citealt{Maozetal2014} and \citealt{MaedaTerada2016}, a summary of several evolutionary routes to SNe Ia  \citealt{TutukovFedorova2007}), {{{{ and a review that was written after this review \citep{Ruiter2020}, }}}} that together cover all relevant aspects of SNe Ia, in the present study I concentrate on the new results from the year 2019. I discuss only scenarios that involve binary stellar systems, and will not touch single star scenarios {{{{ (e.g., \citealt{Clavelli2019, Antoniadisetal2020}; I note that the evolution after the formation of a massive core inside a single AGB star in the scenario of \citealt{Antoniadisetal2020} has some similar ingredients to the core degenerate scenario). }}}}

In an earlier review \citep{Soker2018Rev} I summarised the five binary SN Ia scenarios in a table, which I find to be the most convenient presentation for this review as well. I bring an updated version of that table in Table \ref{tab:Table1}, and I will refer to this table throughout the review (note that in the new table here I split one scenario to two channels). {{{{ The contributions of the scenarios to normal and peculiar SNe Ia are my present estimates, and one should note that other researchers might have largely different estimates, e.g., \citealt{Ruiter2020}. }}}}  
\begin{table*}
\scriptsize
\begin{center}
  \caption{Confronting SN Ia scenarios with observations}
    \begin{tabular}{| p{2.4cm} | p{2.2cm}| p{2.2cm}| p{2.2cm}| p{2.2cm} | p{2.2cm} | p{2.2cm} |}
\hline  
\textbf{{Scenario$^{[{\rm 1,2}]}$}}  & {Core Degenerate \newline (CD)}    & {Double Degenerate} (DD) & {Double Degenerate} (DD-MED)& {Double Detonation} (DDet) & {Single Degenerate} (SD-MED) & {WD-WD collision} (WWC)\\

\hline  
 \textbf{{Channel by MED}}
  & MED built-in. 
  & No MED
  & MED
  & No MED (not allowed)
  & MED (with no MED peculiar SNe Ia)  
  & No MED \\
\hline  
 {$\mathbf{[N_{\rm exp}, N_{\rm sur}, M]}$}
  & $[1,0,M_{\rm Ch}]$ 
  & $[2,0,$sub-$M_{\rm Ch}]$
  & $[1,0,M_{\rm Ch}]$ 
  & $[2,1,$sub-$M_{\rm Ch}]$
  & $[2,1,M_{\rm Ch}]$  
  & $[2,0,$sub-$M_{\rm Ch}]$ \\
\hline  
\hline  

 {Presence of 2 opposite Ears
in some SNR~Ia$^{[{\rm 3}]}$}
  & Explained by the SN inside planetary nebula (SNIP) mechanism.
  & Low mass Ears if jets during merger \citep{TsebrenkoSoker2013}.
  & Requires a short gravitational waves delay time shortly after CEE; unlikely.
  & No Ears are expected for He WD companion.
  & {\textcolor[rgb]{0.8,0.0,0.8}{ OP:$^{[{\rm 4}]}$ Ears by jets from accreting WD \citep{TsebrenkoSoker2013}.}}   
  & No Ears are expected \\

\hline  

 {Spherical SNRs + low polarisations}
  & Expected in all cases.
  & {\textcolor[rgb]{0.98,0.00,0.00}{Cannot explain. }}
  & Expected in all cases. 
  & {\textcolor[rgb]{0.98,0.00,0.00}{Cannot explain. }}
  & Explained in most cases (with MED).   
  & {\textcolor[rgb]{0.98,0.00,0.00}{Cannot explain. }} \\

\hline  

 {$\approx 1M_\odot$ CSM in Keplers SNR }
  & The massive CSM shell might be a PN.
  & {\textcolor[rgb]{0.98,0.00,0.00}{No CSM shell }}
  & Requires a short gravitational waves delay time shortly after CEE; unlikely.
  & {\textcolor[rgb]{0.98,0.00,0.00}{Any CSM is of a much lower mass. }}
  &  {\textcolor[rgb]{0.8,0.0,0.8}{OP:$^{[{\rm 4}]}$ Can be explained by heavy mass loss from an AGB donor. }} 
  & {\textcolor[rgb]{0.98,0.00,0.00}{No CSM shell }} \\
\hline  
 {The need to synthesis $^{55}$Mn and other elements. }
 &  $M_{\rm Ch}$ can do it 
 &  {\textcolor[rgb]{0.98,0.00,0.00}{Not possible }}
 &   $M_{\rm Ch}$ can do it 
 &  {\textcolor[rgb]{0.98,0.00,0.00}{Not possible }}
 &  $M_{\rm Ch}$ can do it    
 & Unlikely \\
\hline  
 {Main
 Scenario
 Predictions}
 & 1. Single WD explodes \newline 2. Massive CSM in some cases (SNIP). \newline 3. $M_{\rm WD} \simeq M_{\rm Ch}$
 & 1. Sufficient WD-WD close binaries \newline2. DTD~$\propto t^{-1}$ \newline 3. $M_{\rm WD} < 1.2 M_\odot$
 &  1. Single WD explodes \newline 2. $M_{\rm WD} \simeq M_{\rm Ch}$
  & 1. A companion survives \newline 2. Asymmetrical explosion \newline 3. $M_{\rm WD} <1.2 M_\odot$
  & 1. A companion survives \newline 2. $M_{\rm WD} \simeq M_{\rm Ch}$ 
  & Asymmetrical explosion  \\
\hline  
 {General
  Strong
 Characteristics}
 & {\textcolor[rgb]{0.00,0.59,0.00}{1. Explains some SN Ia with H-CSM \newline 2. Spherical explosions \newline 3. Many explosions with $M_{\rm WD} \simeq M_{\rm Ch}$ \newline 4. Explains large SNe Ia population shortly after CEE}}
 & {\textcolor[rgb]{0.00,0.59,0.00}{1. Explains very well the delay time distribution (DTD) \newline 2. Ignition easily achieved}}
 & {\textcolor[rgb]{0.00,0.59,0.00}{Explains very well the delay time distribution (DTD) \newline 2. Many explosions with $M_{\rm WD} \simeq M_{\rm Ch}$ \newline 3. Spherical explosions}}
 & {\textcolor[rgb]{0.00,0.59,0.00}{Ignition achieved}}
 & {\textcolor[rgb]{0.00,0.59,0.00}{1. Accreting massive WDs exist \newline 2. Many explosions with $M_{\rm WD} \simeq M_{\rm Ch}$ \newline 3. Spherical explosions}} 
 & {\textcolor[rgb]{0.00,0.59,0.00}{ Ignition easily achieved  }} \\
\hline  
{ Work for future studies }
 & {\textcolor[rgb]{0.8,0.0,0.8}{1. Ignition process\newline 2. Merge during CEE \newline 3. To solidify the claim for $M_{\rm Ch}$ WDs \citep{BearSoker201}  \newline 4. DTD }}
 & {\textcolor[rgb]{0.8,0.0,0.8}{1. To derive spherical explosions  }}
 & {\textcolor[rgb]{0.8,0.0,0.8}{ 1. Ignition process\newline 2. Merge process \newline 3. To solidify the claim for $M_{\rm Ch}$ WDs \citep{BearSoker201} }}
 & {\textcolor[rgb]{0.8,0.0,0.8}{1. To explain the non-detection of helium  \newline 2. To find surviving companions }}
 & {\textcolor[rgb]{0.8,0.0,0.8}{1. Ignition process\newline 2. To explain the DTD and number of SNe Ia \newline 3. To find surviving companions }} 
 & {\textcolor[rgb]{0.8,0.0,0.8}{1. Find a way to account for $> 1 \%$ of normal SNe Ia. \newline 2. Examine which peculiar SNe Ia the WWC might account for. }} \\

\hline  
\hline  
{Contribution to normal SNe Ia$^{[{\rm 5}]}$} 
 & $\approx 20-50 \%$ 
 & $\approx 20-40 \%$ 
 & $\approx 20-40 \%$
 & $\approx 0-10 \%$ 
 & $\approx 0-10 \%$ 
 & $\ll 1 \%$\\
\hline  
{Contribution to peculiar SNe Ia$^{[{\rm 5}]}$}
 & $\approx 0-10 \%$
 & $\approx 30-70 \%$
 & $\approx 0-10 \%$
 & $\approx 10-30 \%$ 
 & $\approx 20-50 \%$ by the SD scenario without MED  
 & $\approx 1 \%$ \\
\hline  

     \end{tabular}
  \label{tab:Table1}\\
\end{center}
\begin{flushleft}
\small Notes:\\
 \small [1] Scenarios for SN Ia by alphabetical order.
 \\
 \small [2] The first three rows are the names and explosion characteristics; $N_{\rm exp}$ is the number of stars at explosion (1 or 2);  $N_{\rm sur}=1$ if a companion survives the explosion and  $N_{\rm sur}=0$ if no companion survives the explosion ; $M_{\rm Ch}$ and sub-$M_{\rm Ch}$,  indicate whether the mass of the exploding WD(s) at explosion is around the Chandrasekhar mass limit or whether it is sub-Chandrasekhar, respectively. 
 \\ 
 \small [3]  The observations in rows 4-10 refer to normal SNe Ia. \\
 \small [4] OP (only peculiar) means that the SD scenario without MED can explain this observation, but not the SD-MED scenario. Therefore, explanation of the particular observation is possible within the SD channels only for peculiar SNe Ia.  \\
 \small [5] Last two rows present my very crude estimates of the contribution of each channel to normal and peculiar SNe Ia, respectively. 
\end{flushleft}
\end{table*}

One can characterise the scenarios by the following properties. (1) Number of stars at explosion, $N_{\rm exp}$, 1 or 2. (2) Whether a companion survives the explosion,  ($N_{\rm sur}=0$ or $1$). (3) Whether the mass of the exploding WD(s) at explosion is around the Chandrasekhar mass limit, $M_{\rm Ch}$, or whether it is sub-$M_{\rm Ch}$. {{{{ I list these characteristics in the third row of Table \ref{tab:Table1}. }}}} To these we can add the partition to channels that have a merger (or accretion) to explosion delay time (MED; \citealt{Soker2018Rev}), and those that do not (second row of Table \ref{tab:Table1}). Specifically, in systems that have MED a merger event or an accretion process bring the WD to a mass of close to $M_{\rm CH}$, and the explosion occurs a time $t_{\rm MED}$ after the WD has reached the mass $M_{\rm WD} \simeq M_{\rm Ch}$, where $t_{\rm MED}$ is much larger than the dynamical time.
  
{{ Both the core degenerate (CD) and the double degenerate (DD) with MED (DD-MED) scenarios have $\mathbf{[N_{\rm exp}, N_{\rm sur}, M]}=[1,0,M_{\rm Ch}]$. The difference between the CD scenario and the DD-MED scenario is that in the DD-MED scenario there is a delay time from the formation of the two WDs to merger that is determined by gravitational waves, while in the CD scenario the merger occurs during the  common envelope evolution (CEE) phase, and so gravitational waves play no role. A big challenge to the DD-MED scenario is that in many cases the merger of two CO WDs does not lead to a SNe Ia with a MED (e.g., \citealt{Wuetal2019}). }}
     
{{  The non-detection of the expected companions inside supernova remnants (SNRs) in the single degenerate (SD) scenario (e.g., \citealt{LiKerzendorfetal2019}; but I note that \citealt{AblimitMaeda2019a} argue that it will be hard to detect the companion for a magnetic WD), the general no detection of hydrogen (e.g., \citealt{ Holmboetal2019}), and/or the very low mass of hydrogen even when rare observations do show hydrogen (e.g., \citealt{Kollmeieretal2019, Prietoetal2020, Dessarteta2020}), suggest that for normal SNe Ia {\it all} cases must have a MED, what originally was termed spin-up/spin-down evolution as rotation keeps the WD from exploding until it loses angular momentum (e.g., \citealt{Piersantietal2003, DiStefanoetal2011, Justham2011, Boshkayevetal2014, Wangetal2014, Benvenutoetal2015}).
For example, \cite{Kuuttilaetal2019} and \cite{GraurWoods2019} strengthen earlier claims that rule out hot and luminous progenitors as expected in some channels of the  SD scenario. Their arguments do not apply to the SD-MED scenario (spin-up/spin-down) that in principle might account for a fraction of normal SNe Ia.
\cite{MengLi2019} suggest that the common-envelope wind channel of the SD-MED scenario \citep{MengPodsiadlowski2017}, can leave a surviving sdB companion which is hard to detect. They further estimate that SNe Ia with an sdB companion might contribute $22 \%$ of all SNe Ia. In the present study I refer to the SD-MED channel when referring to normal SNe Ia, and to the SD scenario without MED when referring to peculiar SNe Ia, unless stated otherwise. For that, I do not list as severe problems in the table the no detection of a surviving companion and the no detection of hydrogen in the ejecta. However, a companion should still survive even if a faint one, and in some cases we should expect hydrogen.
}}
  
     %

I notice that recent results by \cite{Toonenetal2018} solidified  earlier claims  that the WD-WD collision (WWC) scenario  might at most account for less than one per cent of all SNe Ia (later studies have reached the same qualitative conclusion, e.g., \citealt{HaimKatz2018, HallakounMaoz2019, HamersThompson2019}).
The second severe problem of the WWC scenario is that the collision of the two WDs lead to a highly non-spherical explosion (e.g., \citealt{Kushniretal2013}), contrary to both the morphologies of SNRs Ia and to the new polarisation studies of SNe Ia. I do note that there are some studies in 2019 that claim for the WWC scenario (e.g., \citealt{Wygodaetal2019a, Wygodaetal2019b}), and that this scenario might account for rare types of peculiar SNe Ia.
In particular, \cite{Vallelyetal2020} find that some SNe Ia show bimodal velocity distributions of $^{56}$Ni decay products, and \cite{LivnehKatz2020} find evidence for asymmetrical Si distribution. These papers claim that their results support the WWC scenario for many SNe Ia. In the present review I attribute such bimodal and asymmetrical distributions to large $^{56}$Ni and Si clumps in the explosion mechanism, although the global explosion in normal SNe Ia is spherical {{{{ (see, e.g., \citealt{LevanonSoker2019b, Mageeetal2020}, for papers on the $^{56}$Ni distribution). }}}}

 Table \ref{tab:Table1} has several significant differences from the previous comparison table \citep{Soker2018Rev}. 
\begin{enumerate}
\item In my previous review \citep{Soker2018Rev} I used the comparison of the five scenarios with each other and with observations to argue that scenarios must allow, at least in some cases, for a time delay between the dynamical process of merger or the dynamical process of mass accretion to explosion. This merger to explosion delay (MED) was the main conclusion then. In the present study I incorporate the MED as a property that splits scenarios to different channels (second row in Table \ref{tab:Table1}).
\item The outcomes of the DD scenario when explosion occurs within several dynamical time scales after merger or at much later times (the DD-MED) are very different. For that, I find it necessary to split this scenario to two columns in Table \ref{tab:Table1}. A MED might take place when the merger remnant has a mass of $M_{\rm rem} \simeq M_{\rm CH}$.  
\item I changed somewhat the rows. For example, to emphasise the new polarisation studies that suggest spherical explosions, I added the fifth row. 
\item Following the above changes and new results from 2019, I changed my estimates of the contribution of each channel to normal and peculiar SNe Ia.  
\end{enumerate}
 
 
I turn to discuss some new studies from 2018/2019 and their implications to the different scenarios. 
These new studies join older studies (that I do not review here) in challenging one or more scenarios. When comparing theoretical studies to observations these challenges cannot be ignored. There is no scenario that is free from challenges, some that I find hard, or even impossible, to overcome. For example, I do not see how the WWC scenario overcome the two challenges I mentioned above.  

I note again that I cite mainly papers from 2019, although in most cases there are earlier relevant papers, because this review is about studies from 2019 (and to some extend from 2018) and their new implications. 

\textbf{$\bigstar$ Summary of section \ref{sec:intro}.} It is mandatory to consider all scenarios when comparing observations to theory. There are five (5; not 2) scenarios, with some of them having multiple channels. Most significant splitting to channels is that to the two channels of the DD scenario according to the presence (the DD-MED channel) or not of a MED (merger to explosion delay) time. I find the observation that most normal SNe Ia have globally spherical explosions to be one of the most significant in 2019. This finding, alongside others observations, strongly limit the contribution of the DDet scenario, the SD scenario without MED, and the DD scenario without MED to normal SNe Ia.
   
\section{Spherically symmetric explosions}
\label{sec:Spherical}

\subsection{New polarisation observations}
\label{subsec:Polarisation}

I consider the observations of very low polarisation of SNe Ia \citep{Cikotaetal2019, Yangetal2019} to be the most significant type of observations published in 2019. These observations suggest that most SNe Ia have spherical explosions. The globally spherical morphologies of many SNRs Ia (but not all are spherical, e.g.,  \citealt{Alsaberietal2019}) also suggest that most SNe Ia have spherical explosions (review by \citealt{Soker2018Rev}), but the new observations put this conclusion on the forefront (for a summary of polarisation of SNe Ia see, e.g., \citealt{Mengetal2017}). 
I note that \cite{Fangetal2020} show that they can reproduce the departure of the SNR of SN~1006 from spherical morphology by an interaction of a spherical explosion with an ambient medium with a density discontinuity.
As well, \cite{Lukenetal2020} conclude that the young type Ia \citep{Borkowskietal2013} SNR~G1.9+0.3 that has an asymmetrical structure expands into an inhomogeneous interstellar medium (ISM). 

{{{{ Some features, such as `Ears' and other protrusions can exist in otherwise spherical explosions. Some of these are iron-rich protrusions (`iron-bullets'; e.g., \citealt{TsebrenkoSoker2015b, Satoetal2020}).  }}}}

\cite{Cikotaetal2019} study the polarisation of 35 SNe Ia, and argue that their results support the possibility of two distinct explosion mechanisms. Their analysis shows the peak polarisation of the Si~II line to be consistent with the expectation of the DDet scenario and of the delayed-detonation mechanism of $M_{\rm Ch}$ explosions. The violent merger of the DD scenario predicts (e.g., \citealt{Bullaetal2016DD}) too high polarisation (e.g., for 34 out of 35 SNe Ia. 

\cite{Yangetal2019} note that their detection of low polarisation in the normal SN Ia SN~2018gv implies a high degree of spherically symmetric explosion, and that this in turn is consistent with the expected morphology of the delayed detonations explosion mechanism and is inconsistent with the merger-induced explosion mechanism. 

For the benefit of the discussion in section \ref{sec:DelayTimes}, I recall that not all SNe Ia have low polarisation. \cite{Cikotaetal2017} find that the polarisation curves of some SNe Ia sight-lines are similar to those of some proto-planetary nebulae (PNe). They claim that this suggests that some SNe Ia explode inside the wind of a post-asymptotic giant branch star. The case of a SN Ia that explodes inside a proto-PN or a PN is termed SNIP \citep{TsebrenkoSoker2015a}, for a SN inside a PN  \citep{DickelJones1985}.

The findings that most SNe Ia have global spherically symmetric explosions strongly challenges several explosion mechanisms.  \cite{Ferrandetal2019} show in a recent paper that morphological signatures of the explosion can still be detected hundreds of years after explosion, so that an interaction with the ISM cannot erase global asymmetries of SN Ia explosions.

I here discuss only two recent papers on explosion mechanisms. 
But I do note that alongside a global spherical structure of the SN Ia ejecta, the ejecta might be clumpy (e.g., \citealt{Millardetal2019, Satoetal2019}), particularly the iron group elements (e.g., \citealt {Maguireetal2018}).   

\subsection{The DDet scenario}
\label{subsec:DDet}

I consider first the DDet scenario. This scenario has some strong points, e.g., \cite{Townsleyetal2019} perform hydrodynamical simulations of the DDet scenario and find that it can account for the brightness and spectra of SNe Ia (see also, e.g., \citealt{Polinetal2019a}). {{{{ \cite{Bullaetal2020} find the early light curve of many SNe Ia to be compatible with the DDet scenario. }}}}
However, the observations that suggest spherical SN Ia explosions strongly challenge this scenario. I do note that \cite{Bullaetal2016DDet} calculate the polarisation of the DDet scenario and find it to be low enough to be compatible with observations. {{{{ However, \cite{Bullaetal2016DDet} use an explosion model from \cite{Finketal2010} that does not include the roles of the companion (helium-donor) star in causing deviation from a spherical explosion. Namely, the companion forms an empty conical region in the ejecta by blocking the SN ejecta along its direction, and the companion causes a departure from symmetry by its orbital motion (see below). \cite{Bullaetal2016DDet} perform their simulations in 2D by adopting an azimuthal symmetry about the z-axis, rather than in full 3D. For that, it might be that \cite{Bullaetal2016DDet} underestimate the polarisation of the DDet scenario. }}}}

The first process that leads to a non-spherical explosion is the collision of the ejecta with the surviving companion. This collision forms a conical region behind the companion where the density is very low (e.g., \citealt{Tanikawaetal2018, Tanikawaetal2019, Baueretal2019}).  
But there is another process that causes deviation from spherical explosion. 

Consider the ``dynamically driven double-degenerate double-detonation'' (D$^6$) scenario that \cite{Shenetal2018} suggest as an explanation to the three hypervelocity WDs that they identify with Gaia (see simulation by \citealt{Tanikawaetal2019}). Each of this three WDs is the companion that survived a DDet SN Ia. \cite{Neunteufeletal2019} find that the parameter space of the DDet scenario with a non-degenerate helium donor might account for no more that $3 \%$ of all SNe Ia. In the D$^6$ model, though, the helium-donor is a WD.  
 
According to \cite{Shenetal2018}, the respective velocities of the three hypervelocity WDs relative to the Galaxy are $v_2 \simeq 1300 \km \s^{-1}$,  $2300 \km \s^{-1}$, and  $2400 \km \s^{-1}$, with large uncertainties. In the DDet scenario these velocities are the pre-explosion orbital velocities of the respective helium-donor WD companion in each system. To achieve a high velocity of $v_2 \simeq 2000 \km \s^{-1}$ the mass of the WD companion is $M_2 \simeq 1 M_\odot$ \citep{Shenetal2018}. 
Since the exploding WD has a similar mass, the orbital velocity of the exploding WD is $v_{\rm 1,orb} \simeq v_2 \simeq 2300 \km \s^{-1}$ in the two cases above with high hypervelocities. 
What causes departure from spherical explosion is that during the explosion the companion and the ejecta do not move on straight lines, bur rather continue to curve in the sense of their original orbital motion. This implies that the direction of ejection of the center of mass of low-velocity ejecta is not as that of higher velocity ejecta. Let us demonstrate this qualitatively. 

Consider then the explosion process of the WD in the presence of a helium-donor star of similar mass.  For a demonstrative case I take the spherical density profiles from \cite{DwarkadasChevalier1998}
\begin{equation}
    \rho_0(v,t) = \frac{M_e}{8 \pi \left( v_e t \right)^3} e^{-v/v_e}, \quad
    v_e = \left( \frac{E_k}{6M_e} \right)^{1/2},
\label{eq:exponential profile}
\end{equation}
where $M_e$ and $E_k$ are the total ejecta mass and kinetic energy, respectively. For $M_e=1M_\odot$ and $E_k=10^{51} \erg$ we find $v_e=2895 \km \s^{-1}$. Integration over the volume shows that half of the ejecta mass is within a velocities of $v_{\rm h}=7750 \km \s^{-1} \simeq 3.4 v_{\rm 1,orb}$, and quarter of the mass within a velocity of $v_{\rm q}=5000 \km \s^{-1}$. 
By the time the coordinate of half the mass expands beyond the companion, half of the mass is still under the influence of the gravity of the companion, and it changes its direction as if the companion continues to orbit the center of mass more or less. This time is $t_{\rm h} \simeq 2R_{\rm 1,orb}/7750 \km \s^{-1}$, where $R_{\rm 1,orb}$ is the orbit of the exploding WD around the center of mass, which is about equal to that of the surviving companion.  During that time the mass still within the orbit of the secondary star would crudely orbit an angle of 
$\alpha_{\rm h} \approx 360^\circ v_{\rm 1,orb} t_{\rm h} /2 \pi R_{\rm 1,orb} \simeq 360^\circ/3.4 \pi = 34^\circ$.
In the same manner we find that for quarter of the mass $v_{\rm q}=5000 \km \s^{-1} \simeq 2.2 v_{\rm 1,orb}$ and $\alpha_{\rm q} = 53^\circ $

There is a need for more accurate calculations to calculate the polarisation days after explosion, and to reveal the later SNR morphology. But this simple calculation shows that the D$^6$ scenario with a companion that escapes the system after explosion with $v_2 \ga 2000 \km \s^{-1}$ has highly non-spherical explosion, contrary to observatrions. 

In the case of $v_2 \simeq 1300 \km \s^{-1}$ the mass of the hypervelocity WD is $M_2 \approx 0.2 M_\odot$, and the orbital velocity of the WD that exploded was $v_{\rm 1,orb} \simeq 300 \km \s^{-1}$. In that case the orbital motion has a minor influence on the explosion morphology, which will nonetheless be non-spherical due to the collision of the ejecta with the companion.  

The production of a non-spherical SNe Ia is not the only recent problem of the DDet scenario. \cite{Polinetal2019b} show that although the DDet scenario can qualitatively reproduce
sub-luminous SNe Ia spectra in the nebular phase, these explosions produce too much [CaII] emission compared to most normal SNe Ia. 

Another general problem is that sub-$M_{\rm Ch}$ explosions cannot yield some isotopes (e.g., \citealt{Bravo2019}), such as $^{55}$Mn (seventh row in Table \ref{tab:Table1}). 
         
\subsection{The hybrid HeCO channel of the DD scenario}
\label{subsec:Hybrid}

The second study I consider deals with the hybrid HeCO channel of the DD scenario  \citep{Zenatietal2019, Peretsetal2020}. (The HeCO hybrid channel of the DD scenario is different than the hybrid CONe channel of $M_{\rm Ch}$ deflagration to detonation explosions, e.g., \citealt{Augustineetal2019}).
\cite{Peretsetal2020} simulate the destruction of a HeCO WD by a CO WD with 2D hydrodynamical simulations that include nuclear reactions. \cite{Peretsetal2020} set their initial conditions to have a HeCO torus around the CO WD. 
They obtain a double detonation process and ignite both the CO WD and the HeCO torus. They can account for several important properties of SNe Ia, like the light curves, spectra,  and the range of peak-luminosities. They further argue that together with the contribution from the DD scenario of two massive CO WDs they can reproduce the rate and delay-time distribution of SNe Ia.

However, they face several challenges. Their 2D code forces them to simulate axisymmetrical flow.
For that, they could not actually follow the destruction of the HeCO WD and establish whether the destructed HeCO WD forms a torus before ignition. The second challenge is the high density along the symmetry axis inside the CO WD that they obtain in their simulations {{{{ (their figure 1). }}}} Their 2D axisymmetrical numerical grid necessarily forces the shock wave that the ignition of the HeCO torus induces in the CO WD to converge on, or near, the axis {{{{ (their figure 1 at $t=7.2 \s$). }}}} Namely, their 2D grid ignites a ring around the symmetry axis, rather than a point in reality. In reality, the ignition that starts in one point will not lead to an axisymmetrical shock wave, {{{{ but rather will converge on an off-center point (e.g., \citealt{Pakmoretal2013, Townsleyetal2019, Gronowetal2020}). }}}} It is not clear whether \cite{Peretsetal2020} can obtain high enough densities inside the CO WD. I also note that even with these high densities (which night be overestimated) they do not synthesis some elements, such as $^{55}$Mn.    
       
I consider the third challenge to be the strongest one. \cite{Peretsetal2020} present their numerical grid up to a distance of $6 \times 10^9 \cm$ from the center of the flow. From what they present {{{{ (their figure 1), }}}} one sees that the explosion is highly non-spherical. There is a fast polar outflow and much slower equatorial ejecta. It is a challenge for the hybrid HeCO channel of the DD scenario to show they can account for the new low polarisation measurements \citep{Cikotaetal2019, Yangetal2019}, and for globally spherical SNRs Ia {{{{ (section \ref{subsec:Polarisation}). }}}}   
   
\textbf{$\bigstar$ Summary of section \ref{sec:Spherical}.} The new polarisation studies from 2019 \citep{Cikotaetal2019, Yangetal2019} impose firm constraints on explosion mechanisms to yield explosions that are globally spherical, at least in most cases, if not in all normal SNe Ia. (Small features, such as `Ears' are possible; Table \ref{tab:Table1}.) Interestingly, interactions that seem more easily to explode in numerical simulations have highly non-spherical ejecta. These include the violent merger of the DD scenario, the WWC scenario, possibly the D$^{6}$ channel of the DDet scenario, and the newly simulated hybrid HeCO channel of the DD scenario, as I discussed above. The deflagration-to-detonation explosion mechanism of $M_{\rm Ch}$ WDs lead to spherical explosions, and so I favour this explosion mechanism. Although there is a need to work out several steps in this explosion mechanism, the new study by \cite{Poludnenkoetal2019} strongly supports the case for the deflagration-to-detonation explosion mechanism.

\section{Early excess emission}
\label{sec:EarlyEmission}

An interesting class of SNe Ia is that of SNe Ia that show early ($\la 5 \days$) excess emission in their light curve (e.g., \citealt{Jiang2018, LiWang2019, Shappee2019, Dimitriadisetal2019a}). 
{{{{ In this section I use the class of SNe with early excess emission to emphasise that in many cases we cannot attribute a given type of observations (here it is the early excess emission) to only one scenario (like the SD scenario). We must examine the possibility that other SN Ia scenarios might also account for that specific type of observations. }}}}
  
The SD scenario predicts this kind of early emission in most SNe Ia (e.g. \citealt{Kasen2010}), unless there is a very long MED during which the mass-donor radius decrease by an order of magnitude or more, i.e., a giant donor that becomes a WD. My point here is that  such an emission is possible also in the DD scenario if the explosion takes place within a time of about $10^3 \s \la t_{\rm MED} \la 1~$day, as the ejecta collides with disk-originated matter (DOM; \citealt{LevanonSoker2019}).  
    
The positive side for the DD scenario of the requirement of $t_{\rm MED} \ga 10^3 \s \gg t_{\rm dyn}$, where $t_{\rm dyn} \simeq 10 \s$ is the dynamical time of the merger process, is that the merger product has time to acquire spherical structure that will lead to the required spherical explosion (section \ref{subsec:Polarisation}).  
The rarity of the early excess emission might suggest that in most cases the MED of the DD scenario is longer than tens of years to allow the DOM to disperse \citep{Levanonetal2015}, again, leading to spherical explosion. 

The finding that early excess emission is rare (e.g., \citealt{Fausnaughetal2019}), and the non-detection of hydrogen in the ejecta (e.g., \citealt{Dimitriadisetal2019b}), or only a very low mass of hydrogen in the ejecta (section \ref{sec:Hydrogen}), imply that if the SD scenario accounts for some SNe Ia, there must be a very long MED. Namely, it must be the SD-MED scenario. 
 
\textbf{$\bigstar$ Summary of section \ref{sec:EarlyEmission}.} Early excess emission can occur in principle both in the SD and in the DD scenarios. The new findings of early excess emission and their rarity suggest that in both scenarios the common channel is the one where there is a merger (or accretion) to explosion delay (MED) time. Namely, for normal SNe Ia the DD-MED and SD-MED scenarios dominate the DD and the SD scenarios, respectively.  

\section{Hydrogen in SNe Ia}
\label{sec:Hydrogen}

Most recent studies find no hydrogen in SNe Ia (e.g., \citealt{Tuckeretal2020}). This by itself almost rules-out the SD scenario that has no MED. It leaves  the SD-MED channel (spin-up/spin-down channel) for normal SNe Ia. But there is still the problem that population synthesis studies cannot attribute all SNe Ia to the SD scenario. 

Even the recent detection of hydrogen in two SNe Ia is much below that expected in the SD scenario. 
\cite{Kollmeieretal2019} estimate the hydrogen mass in the sub-luminous SN Ia SN2018fhw/ASASSN-18tb to be $M_{\rm H} \approx 2 \times 10^{-3} M_\odot$ (or maximum up to $ M_{\rm H} \approx 0.01 M_\odot$).  
To explain the H$\alpha$ line they consider stripped gas from the companion in the SD scenario, fluorescent UV pumping in a slowly expanding shell of material, and interaction with CSM (also \citealt{Vallelyetal2019a}). However, \cite{Vallelyetal2019a} note that SN2018fhw is very different from other known CSM-interacting SNe Ia. 

\cite{Prietoetal2020} infer a hydrogen mass of $M_{\rm H} \approx 10^{-3} M_\odot$ from their detection of H$\alpha$ emission line in the low-luminosity fast declining SN Ia SN2018cqj/ATLAS18qtd. Although some properties of the H$\alpha$ line are consistent with stripped hydrogen, the hydrogen mass is significantly less than predictions of theoretical calculations of the SD scenario without MED (e.g., \citealt{Botynszkietal2018}).

Another explanation for the presence of little hydrogen, although speculative, is the rare evaporation of a planet (or two) by SNe Ia \citep{Soker2019planet}. The Jupiter-like planet should reside within $a \simeq 50 R_\odot$ from the SN. Interestingly, all SN Ia binary scenarios might in principle account for the presence of planets.

\textbf{$\bigstar$ Summary of section \ref{sec:Hydrogen}.} The general non-detection of hydrogen in SNe Ia, and the very low hydrogen mass of $M_{\rm H} \approx  10^{-3} M_\odot$ in rare cases, rules out the SD scenario that has no long MED ($t_{\rm MED} \ga 10^7 \yr$; \citealt{Soker2018Rev}). 
The very low hydrogen mass in rare cases might come from a planet that the SN Ia evaporates. 

\section{Delay times and their implications}
\label{sec:DelayTimes}

The delay time distribution (DTD) refers to the distribution of SNe Ia with time from star formation to explosion, $t_{\rm SF-E}$. A common approximation is a power-law {{{{ (see \citealt{Strolgeretal2020} for an exponential form)  }}}}
\begin{equation}
\dot N_{\rm DTD} \equiv \left( \frac {d N_{\rm Ia}}{dt} \right)_{\rm DTD} = A \left( \frac{t}{1 \Gyr} \right)^{\alpha}. 
\label{eq:dotN}
\end{equation}
In the past, many studies have obtained $\alpha \simeq -1$. Two significant recent papers find steeper dependence. \cite{FriedmannMaoz2018} derive $A = 5-8 \times 10^{-13} M^{-1}_\odot \yr^{-1}$ and $\alpha=-1.30^{+0.23}_{-0.16}$, and \cite{Heringeretal2019} derive $A = 7\pm 2 \times 10^{-13} M^{-1}_\odot \yr^{-1}$ and $\alpha=-1.34^{+0.19}_{-0.17}$, although other derivation exist (e.g., \citealt{Frohmaieretal2019}).  
 
The time $t_{\rm SF-E}$ can be composed from several times. For scenarios that involve two WDs there are in principle 3 times. These are the time from star formation to the formation of the two WDs in the post-CEE phase, $t_{\rm SF-CE}$, the time for gravitational waves to merge the two WDs, $t_{\rm GW}$, and the time from the merger of the two WDs to explosion, the MED time $t_{\rm MED}$.  Namely, 
\begin{equation}
t_{\rm SF-E}({\rm DD}) = t_{\rm SF-CE} + t_{\rm CEED} = t_{\rm SF-CE} + t_{\rm GW} + t_{\rm MED} , 
\label{eq:DDtSFE}
\end{equation}
where $t_{\rm CEED}=t_{\rm GW} + t_{\rm MED}$ is the time from the end of the CEE to explosion. 
In the DD and DD-MED scenario the longest time is $t_{\rm GW}$. In the DD scenario and in the DDet scenario with a WD helium-donor, such as the D$^6$ channel, $t_{\rm MED}=0$. In the CD scenario the merger takes place during the CEE and so $t_{\rm GW}=0$ and $t_{\rm CEED} = t_{\rm MED}$. 

There are some SNe Ia that occur while there is a CSM around the explosion cite, so called SNe-CSM. 
\cite{Grahametal2019} estimate the fraction of SNe Ia-CSM with close CSM to be $f_{\rm CSM} < 0.06$ of all SNe Ia. In addition, there are cases with CSM further away. 
Over all, the presence of a CSM implies that in scenarios that involve a CEE the explosion occurs within a short time after the CE, $t_{\rm CEED} \la 10^6 \yr$. {{{{ Of course, in most cases explosions occur much later, e.g., as non-detection of radio emission reveals (e.g., \citealt{Cendesetal2020, Lundqvistetal2020}).  }}}}

I used the new findings of \cite{Grahametal2019} and of \cite{FriedmannMaoz2018} and \cite{Heringeretal2019}, to study the relation of SNe Ia that occur at short times (within a million years) after the CEE, and those that take place long after the CEE. 
{{{{ Based on these studies that show from observations this population of SNe Ia to be young \citep{Grahametal2019},  }}}} I concluded \citep{Soker2019CEE} that the population of SNe Ia with $t_{\rm CEED} \la 10^6 \yr$ amounts to $\approx {\rm few} \times 0.1$ of all SNe Ia. I further concluded that the expression for the rate of these SNe Ia is different from that of the DTD billions of years after star formation (equation \ref{eq:dotN}). 
{{{{ In a recent study \cite{Jerkstrandetal2020} suggest that the superluminous SN~2006gy exploded in the frame of the CD scenario shortly after the CEE. Their suggestion, if holds, supports my earlier conclusion on the importance of the SN Ia population with short $t_{\rm CEED}$. }}}}

From these conclusions I suggested that the physical processes that determine the explosion time shortly after the CEE are different (at least to some extend) than the processes that determine the explosion times long after the CEE. I here argue that at least those that occur within a million years after the CEE come from the CD scenario.
I do note thought that there are claims that the CD scenario contributes less than my estimate in the Table \ref{tab:Table1} (e.g., \citealt{Wangetal2017C}). 

\textbf{$\bigstar$ Summary of section \ref{sec:DelayTimes}.} New results from 2018-2019 strengthen the case for a population of SNe Ia that occur within a million years after the CEE of their progenitors. The other population is of SNe Ia that occur much later, over a timescale of $\approx 10^7 \yr$ and longer. 
The processes that determine the delay time from the CEE to explosion are different (at least somewhat) in these two populations. I suggest that the short-delay population comes mainly from the CD scenario. 

\section{Peculiar SNe Ia}
\label{sec:Peculiar}
   
Peculiar SNe Ia include SNe Iax, Ca-rich transients, SN~2002es-like SNe Ia,  SN~1991bg-like SNe Ia, 2009dc-like SNe Ia-pec, and more (e.g., review by \citealt{Jhaetal2019NatAs}). \cite{Chenetal2019ApJ880} study a luminous SNe Ia, and suggest to replace the name ``Super-Chandrasekhar SN Ia'' by ``2009dc-like SN Ia-pec'', that I use here. In Table \ref{tab:Table1} I list my crude estimate of the contribution of different scenarios to peculiar SNe Ia. (\citealt{MaedaTerada2016} present a different type of table of some scenarios and their properties in relation to normal and peculiar SNe Ia but that table does not include the CD scenario nor estimates of the fractional contribution of each scenario.) 

In the summary of section \ref{sec:Spherical} I noted that dynamical binary interactions that numerical simulations show to ignite one or both WDs (e.g., the violent merger,  the D$^{6}$, the hybrid HeCO channels of the DD scenario, and the WWC scenario), have highly non-spherical explosions. Although some studies show that the ignition in the DDet scenario might not be easy (e.g., \citealt{WuWnag2019}), I assume that ignition does take place (e.g., \citealt{Gronowetal2020}) but lead to highly asymmetrical explosions (section \ref{subsec:DDet}). For that, I find that these scenarios and channels are unlikely to account for a large fraction of normal SNe Ia. However, they might explain a large fraction of peculiar SNe Ia and other transients (e.g., like accretion-induced collapse, e.g., \citealt{Wang2018AIC}), and the easy ignition in these processes suggests that peculiar SNe Ia (and similar transients) are very common, possibly even more common that normal SNe Ia. Not surprisingly, there are relatively many recent papers on peculiar SNe Ia and similar transients (e.g., \citealt{LyutikovToonen2019, Pantheretal2019, Prenticeetal2020}).

The studies of peculiar SNe Ia from 2019 add to the growing evidence that peculiar SNe Ia come from several channels and scenarios, possibly from all scenarios listed in Table \ref{tab:Table1}. Indeed, at least some peculiar SNe Ia show low polarisation as SNe Ia show \citep{Mengetal2017}. 
     
The DDet scenario seems to be popular by studies of peculiar SNe Ia (e.g., \citealt{Mageeetal2019}), {{{{{ and in particular Ca-rich transients (e.g., \citealt{DeKasliwaletal2020}). }}}}} \cite{Polinetal2019b} study the DDet scenario and argue that even a small amount of calcium can make a Ca-rich transient. \cite{JacobsonGalanetal2019} argue that the Ca-rich SN~2016hnk is consistent with the DDet scenario. \cite{Takaroetal2020} estimate the delay time from star formation to explosion of SNe Iax, and conclude that they better fit the DDet scenario than massive WR progenitors. 
However, other scenarios exist for these types of peculiar SNe Ia. For example, the hybrid HeCO channel of the DD scenario for Ca-rich transients \citep{Zenatietal2020}.  

\cite{Sandetal2019} find no H$\alpha$ emission in fast-declining SNe Ia, suggesting these do not come from the SD scenario. But not all peculiar SNe Ia are sub-$M_{\rm Ch}$ explosions. \cite {Galbanyetal2019} argue that the peculiar SN~2016hnk, a Ca-strong 1991bg-like SN, comes from an $M_{\rm Ch}$ explosion. 
\cite{JacobsonGalanetal2019MN} spectroscopically model 44 SNe Iax. They find helium in two SNe Iax that better fit a CSM helium than a helium in the ejecta. In the majority of SNe Iax they find no helium, but still suggest that the SD scenario with helium donor might be the main contribute of SNe Iax. 
\cite{MengPodsiadlowski2018Pec} propose that SNe Iax that are SNe Ia-CSM result from the hybrid CONe common-envelope wind channel of the SD scenario. 

\cite{Raddietal2019} report the observation of three runaway stars, that are chemically peculiar due to enrichment of nuclear ashes of partial oxygen and silicon burning. They deduce their masses to be in the range of $0.2-0.28 M_\odot$.  
They further speculate that these inflated WDs are the partly
burnt remnants of either electron-capture supernovae or peculiar SNe Iax (e.g., \citealt{Zhangetal2019}).
My view is that \textit{any} thermonuclear explosion that leaves a runway companion, e.g., simulations by \cite{Baueretal2019} of the DDet (D$^6$) scenario, belongs to peculiar SNe Ia (and other transients), such as SNe Iax. 

\textbf{$\bigstar$ Summary of section \ref{sec:Peculiar}}.
The relative easy ignition of WDs by dynamical effects in numerical simulations of some scenarios and channels (e.g., accretion at a high rate or collision), as evidence also from studies from 2019, suggests that these explosions are common. However, since they are highly globally non-spherical, they cannot explain many normal SNe Ia (section \ref{sec:Spherical}). These non-spherical explosions, I suggest, make the majority of peculiar SNe Ia, as I summarise in Table \ref{tab:Table1}. 
I expect, therefore, that most peculiar SNe Ia have non-spherical explosions, unlike normal SNe Ia. 

\section{SUMMARY}
\label{sec:summary}

The question of the main progenitors of normal SNe Ia is unsolved yet. Not only there is no consensus on the most promising SN Ia scenario(s), but there is no consensus even on the different potential scenarios. The correct counting as of the end of 2019 is of five binary scenarios: the CD, the DD, the DDet, the SD and the WWC scenario (section \ref{sec:intro}). 
I list them in Table \ref{tab:Table1}, where I also split the DD scenario to a channel without a merger to explosion delay (MED) time, and channel with a MED, the DD-MED channel (scenario).
(It might be that we need to consider the DD channel and the DD-MED channel of the DD scenarios as two separate scenarios.) 

According to population synthesis studies there are not enough double WD systems with combined mass of $\ga M_{\rm Ch}$ (e.g., \citealt{Chengetal2020}; I note that \citealt{RebassaMansergasetal2019} argue that observationally it is not easy to detect WD binary systems). Therefore, the DD-MED scenario cannot account for all SNe Ia. 

The strongest point of sub-$M_{\rm Ch}$ scenarios (DD, DDet, WWC scenarios) is that they can achieve easy ignition by dynamical processes. However, they lead to highly non-spherical explosions, in contradiction with observations (Table \ref{tab:Table1}). On the other hand, the main challenge of the $M_{\rm Ch}$ scenarios (CD, DD-Med and SD scenarios) is to achieve ignition. 
Recent studies (e.g., \citealt{Fisheretal2019, Poludnenkoetal2019}) show that turbulence can drive detonation (deflagration-to-detonation) in WDs, and by that turbulence increases the range of conditions for the onset of carbon detonation. This result helps the $M_{\rm Ch}$ explosions. 
On the observational side, \cite{Kawabata2019} study the high-velocity SNe Ia SN~2019ein, and find its properties to be compatible with deflagration to detonation $M_{\rm Ch}$ explosion. 
However, there are still open issues and observations that the $M_{\rm Ch}$ model cannot explain (yet), e.g., \cite{Byrohletal2019}.   
 
The entire set of observations brings many researchers to the conclusion that SNe Ia come both from $M_{\rm Ch}$ explosions and from sub-$M_{\rm CH}$ explosions (e.g., \citealt{LevanonSoker2019b, Polinetal2019a, Polinetal2019b, Sarbadhicaryetal2019, Seitenzahletal2019}). The new study by \cite{Seitenzahletal2019} which presents a new approach to study SNRs is a good example for that. By spatially resolving some nebular emission lines, they effectively perform a "tomography", i.e., they reveal the location of the reverse shock. By using evolutionary model they could constrain the type of explosion in two SNRs. They conclude that SNR~0519-69.0 was a standard $M_{\rm Ch}$ explosion, while SNR 0509-67.5 was an energetic sub-$M_{\rm Ch}$ explosion. As another example, \cite{Ashalletal2019a} support $M_{\rm Ch}$ explosion for most SNe Ia, but note that the subluminous SN~2015bo might come from the merger of two WDs. They claim that this demonstrates the diversity of  explosion scenarios of faint SNe Ia. 
\cite{Kobayashietal2019} study galactic chemical evolution from which they conclude that sub-$M_{\rm Ch}$ Sne Ia contribute up to $25\%$ of SNe Ia in the solar neighbourhood. In dwarf spheroidal galaxies sub-$M_{\rm Ch}$ SNe Ia contribute a higher fraction when star formation took place. \cite{Kirbyetal2019} reach similar conclusions, i.e., that sub-$M_{\rm Ch}$ explosions dominated in dwarf galaxies when they formed stars in the past. At present, $M_{\rm Ch}$ SNe Ia might dominate, as they do in the Milky Way and in dwarf galaxies with extended star formation. These studies of chemical evolution, and the very recent study by \cite{delosReyesetal2020}, also point to both $M_{\rm Ch}$ and sub-$M_{\rm Ch}$ explosions.      
 
\textbf{$\bigstar$ Summary of section \ref{sec:summary}.}
The summary of this section, and in turn of the entire review, are the last two rows of Table \ref{tab:Table1} where I list my crude estimates of the contribution of the different scenarios. This table presents my view that normal SNe Ia come both from $M_{\rm Ch}$ explosions, mainly the CD and DD-MED scenarios, and from sub-$M_{\rm CH}$ explosions, mainly the DD scenario. There is a possible small contribution from the SD and DDet scenarios. 
Peculiar SNe Ia come mainly from the DD, DDet, and SD scenarios, all without MED. 

\section*{Acknowledgements}

{{{{ I thank an anonymous referee for detailed and very useful comments. I thank Mattia Bulla, Keiichi Maeda, Paolo Mazzali, Michael Tucker, Ivo Seitenzahl, }}}} and Xiangcun Meng for useful comments. 
This research was supported by the Israel Science Foundation, by the Prof. Amnon Pazy Research Foundation, and by the E. and J. Bishop Research Fund at the Technion. I presented an earlier version of this review at the Progenitors of Type Ia Supernovae meeting, Lijiang, China, August 05-09, 2019.
 

\label{lastpage}
\end{document}